\documentclass[review]{elsarticle}

\usepackage{hyperref}
\usepackage{amsmath}
\usepackage{latexsym}
\usepackage{amssymb}
\usepackage{graphics}
\usepackage{graphicx}
\usepackage{adjustbox}

\journal{Chaos, Solitons \& Fractals}

\begin{document}

\def\bc{\begin{center}} 
\def\ec{\end{center}}
\def\bea{\begin{eqnarray}}
\def\eea{\end{eqnarray}}
\newcommand{\avg}[1]{\langle{#1}\rangle}
\newcommand{\ket}[1]{\left |{#1}\right \rangle}
\begin{frontmatter}

\title{Beyond the clustering coefficient:\\ A topological analysis of node neighbourhoods in complex networks}

\author[mymainaddress]{Alexander P. Kartun-Giles}
\ead{alexanderkartungiles@gmail.com}

\author[mymainaddress,mysecondaryaddress]{Ginestra Bianconi}

\cortext[mycorrespondingauthor]{Alexander P. Kartun-Giles}
\ead{alexanderkartungiles@gmail.com}

\address[mymainaddress]{School of Mathematical Sciences, \\Queen Mary University of London, London, United Kingdom}
\address[mysecondaryaddress]{The Alan Turing Institute, The British Library, London, United Kingdom}

\begin{abstract}
In Network Science node neighbourhoods, also called ego-centered networks have attracted large attention. In particular the clustering coefficient has been  extensively  used to measure their local cohesiveness. In this paper, we show how, given two nodes  with the same clustering coefficient, the topology of their neighbourhoods can be significantly different, which demonstrates the need to go beyond this simple characterization. We perform a large scale statistical analysis of the topology of node neighbourhoods of real networks by first constructing their clique complexes, and then computing their Betti numbers.  We are able to show significant differences between the topology of node neighbourhoods of real networks and the stochastic topology of null models of random simplicial complexes revealing local organisation principles of the node neighbourhoods. Moreover we observe that a large scale statistical  analysis of the topological properties of  node neighbourhoods is able to clearly discriminate between power-law networks, and planar road networks.
\end{abstract}


\end{frontmatter}

\section{Introduction} 

There has been significant recent interest in the topology and the geometry of networks \cite{Perspectives,Lambiotte,Leskovec_science,giusti}. Network topology and geometry allow us to tackle fundamental theoretical questions concerning the principles determining the emergence of network geometry \cite{Emergent, Hyperbolic} or the proper definition of curvature in the discrete setting of networks and simplicial complexes \cite{Jost1,Jost2,Loll}.
Moreover, network topology has  been shown to be an important tool for network inference, and has been extensively used to explore network structure \cite{Vaccarino1,petri2014homological,maletic,scolamiero,sporns,expert} and dynamics \cite{Ziff,Petri1,Petri2}. 
In this paper, we propose a new  network topology framework to investigate the local structure of real networks determined by the neighbourhoods of their nodes, which we often refer to as \textit{node neighbourhoods}. This approach will allow us to detect non-random statistical properties (with respect to a specific random model) in the local organization of these structures.

Node neighbourhoods, also called \textit{ego-centered networks},  have been studied extensively in Network Science \cite{SW,Newman}. In particular the clustering coefficient of a node is the most famous measure that quantifies the local density of triangles, or the so called \textit{transitivity}, of connections \cite{SW}.
In particular the clustering coefficient has been used extensively to quantify to what extent a network satisfies the principle of \textit{triadic closure}. This principle was originally formulated in the context of social networks, where it is observed that two friends of a common person are more likely to be friends of each other than in a set of random relations. However this tendency of real networks of displaying a large number of  triangles has also been observed in other complex networks. Interestingly it is to mention that triadic closure is a basic  mechanism for generating self-organised communities, as demonstrated by models enforcing this mechanism of network evolution \cite{Santo}.

Recently, higher-order clustering coefficients have been formulated in order to measure the density of cliques larger than triangles in a given node neighbourhood. The closure of higher dimensional cliques has also been used for link prediction \cite{Cannistraci_common,Kleinberg}.
However, the clustering coefficient and its generalisations are insufficient to  fully characterise  the topology of a node neighbourhood, and therefore it is important to develop new Topological Data Analysis tools \cite{Nanda,Skraba} that allow us to go beyond these simple metrics.

 To perform a topological analysis of real networks, the first step is to construct a \textit{simplicial complex} starting from the network. A simplicial complex is a higher-order network structure that is not just formed by nodes ($0$-dimensional simplices) and links ($1$-dimensional simplices) like a network but it is also formed by higher dimensional simplices such as triangles ($2$-dimensional simplices), tetrahedra  ($3$-dimensional simplices), and so on. Starting from a real network, we can extract a simplicial complex (called clique complex) by associating to each $c$-clique of the network a $(c-1)$-dimensional simplex. The topology of the clique complex can be investigated by calculating the Betti numbers, therein measuring the number of connected components (Betti number $\beta_0$), the number of cycles that form $1$-dimensional holes (Betti number $\beta_1$), the number of $2$-dimensional cavities (Betti number $\beta_2$), and their higher dimensional generalisations.

Here we propose a novel topological approach for analysing node neighbourhoods which goes beyond the traditional measure of the clustering coefficient, performing a large scale statistical analysis of the topology of the node neighbourhoods of real network datasets, with size ranging from 82,168 to 12,394,385 nodes. The results we obtain are compared to random simplicial complexes \cite{kahle2009} or random Vietoris-Rips complexes \cite{kahle2}, which take on the role of null models. We show how the topology the node neighbourhoods of real datasets significantly differ from these null models, showing that they obey organisation principles. Moreover, we show how the proposed topological analysis of node neighbourhoods reveals significant statistical differences between scale-free networks, and planar road networks.

The paper is organised as follows. In Sec. \ref{sec:2}  we define networks, simplicial complexes and clique complexes, and discuss how the clique complexes can be extracted from a network dataset. In Sec. \ref{sec:3} we define node neighbourhoods, and we describe their topology in terms of the total number of nodes, link density, and Betti numbers. In Sec. \ref{sec:4} we  show evidence of the relevant diversity found in the topology of real network neighbourhoods with comparable node and link density. In Sec. \ref{sec:5} we study the topology of null model of node neighbourhoods, and we discuss general differences observed between real network neighbourhoods and the null models. In Sec. \ref{sec:6}  we provide the results of a large scale statistical comparison between the topology of neighbourhoods of scale-free hierarchical networks and neighbourhoods of road networks, and we compare the statistical properties of these neighbourhoods with the statistical properties of the considered null models. Finally in Sec. \ref{sec:6} we provide the conclusions.
 
\section{Networks, simplicial complexes and clique complexes}\label{sec:2}
 
 A network is a graph $G=(V,E)$ formed by a set of nodes $V$ and a set of links $E$ that represent the elements of a complex system and their interactions, respectively.
 Networks are ubiquitous and include systems as different as the WWW (web graphs), infrastructures (as airport networks or road networks) and biological networks (as the brain of the protein interaction network in the cell).
 Simplicial complexes represent higher-order networks, which include interactions between two or more nodes, described by simplices. A simplex $\mu$ of dimension $c-1$ is formed by a subset of $c$ nodes. for instance a node is a $0$ dimensional simplex, a link is a $1$-dimensional simplex, a triangle is a $2$-dimensional simplex and a tetrahedra is a $4$-dimensional simplex and so on. A simplicial complex ${\mathcal K}$ is  formed a by a set of simplices that satisfy the following two conditions:
 \begin{itemize}
 \item[(a)] if a simplex $\mu$ belongs to the simplicial complex, i.e. $\mu\in {\mathcal K}$ then any simplex $\mu^{\prime}$ formed by a subset of its nodes is also included in the simplicial complex, i.e. if $\mu^{\prime}\subset \mu$ then $\mu^{\prime}\in \mathcal K$;
 \item[(b)] given two simplices of the simplicial complex $\mu\in {\mathcal K}$ and $\mu^{\prime}\in {\mathcal K}$ then either their intersection belongs to the simplicial complex, i.e. $\mu\cap \mu^{\prime}\in {\mathcal K}$ or their intersection is a null set, i.e. $\mu\cap \mu^{\prime}=\emptyset$.
 \end{itemize}
 Given a simplicial complex it is always possible to extract a network known as the $1$-skeleton of the simplicial complex by considering exclusively the nodes and links belonging to the simplicial complex.
 Conversely, given a network, it is possible to derive deterministically a simplicial complex defining its \textit{clique complex}, obtained by taking a $(c-1)$ dimensional simplex for every $c$-clique in the network. The clique complex is a simplicial complex. In fact, if a simplex is included in a clique complex, then all its sub-simplicies are also included. Moreover any two simplices of the clique complex have an intersection that is either the null set or it is a simplex of the clique complex.

\section{Node neighbourhoods}\label{sec:3}
\subsection{Definition}
\begin{figure*}[!t]
  \begin{centering}
     \begin{adjustbox}{width=\columnwidth}
     \centering
      \includegraphics[scale=0.56]{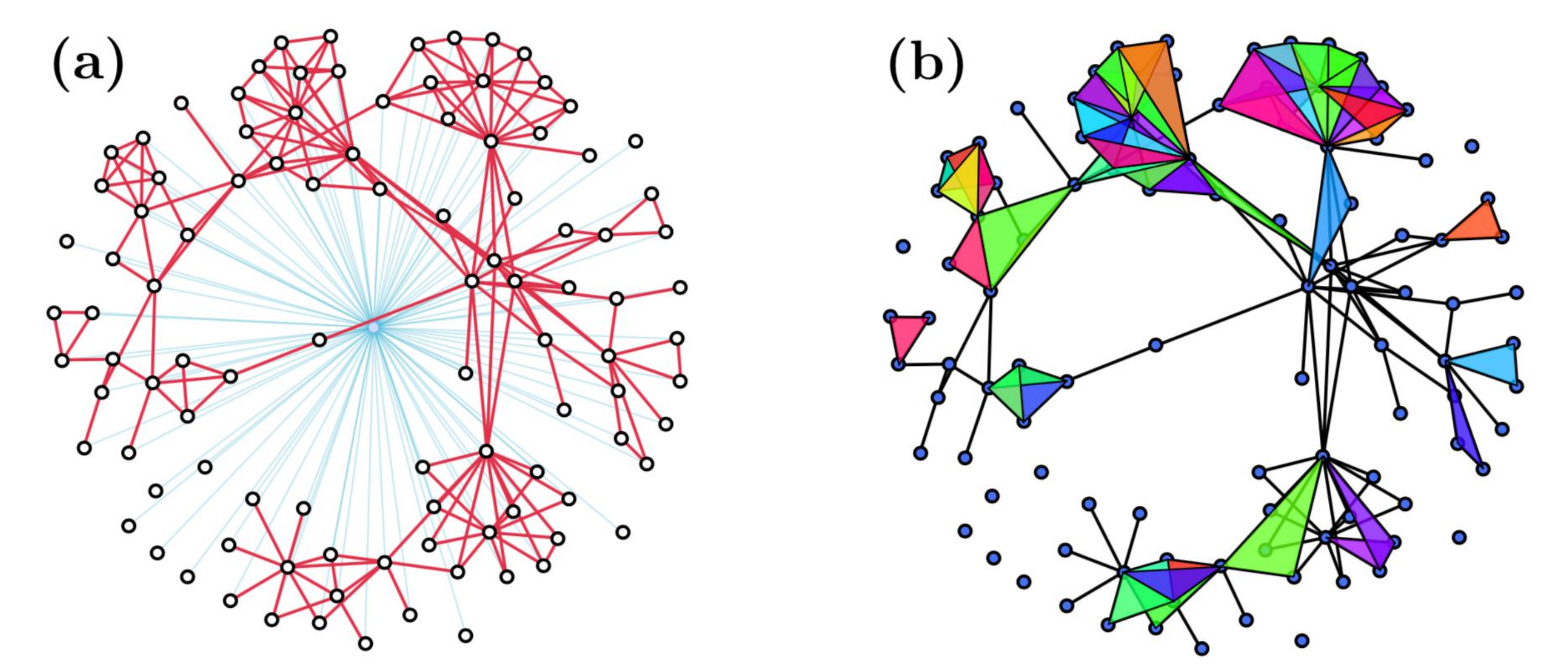}
       \end{adjustbox}
  \caption{The construction of the  neighbourhoods of a node include two steps.  First the subgraph induced by the  neighbours of a node  found at distance $0<\delta\leq d$ is considered (panel (a)). Subsequently the clique complex of this subgraph is constructed, by adding $(c-1)$-dimensional simplices for any  $c$-cliques (panel (b)). }
  \end{centering}
\label{fig:simplicial_complexes}
\end{figure*}

A complex network can be described locally in terms of $d$-hop \textit{neighbourhoods}, or \textit{ego-centered} networks.
Starting from a given node $i$, we consider the subgraph induced by the set of the nodes at hopping distance $\delta$ (with $0<\delta\leq d$). The  neighbourhood of node $i$ is the clique complex of this induced subgraph. For example, if node $i$ has degree 12, of which all nodes apart from one pair are disconnected, the corresponding clique complex contains 11 disconnected components, of which 10 are $0$-simplices, and the other is a $1$-simplex. Fig. \ref{fig:simplicial_complexes} depicts an example of a $1$-hop neighbourhood. On panel (a), the induced subgraph on the neighbours is shown, with the corresponding clique complex on the right in panel (b). The simplices, show up to dimension two, are randomly coloured.

\subsection{Number of nodes  and link density of the neighbourhoods}

The neighbourhood of node $i$ is characterised by its number of nodes $n_i$, and the density of links $\rho_i$, given by 
\bea
\rho_i=\frac{\mbox{\#\ of\ links\ in\ the\ neighbourhood}}{n_i(n_i-1)/2}.
\eea
When the neighbourhoods are formed by nodes at distance $\delta=1$ (i.e.  when we consider $d=1$ and $1$-hop neighbourhoods), the number of nodes $n_i$ is given by the degree $k_i$ of node $i$ in the original network, i.e.
\bea
n_i=k_i.
\eea Additionally in this case  the density of links $\rho_i$ in the neighbourhood is given by the local clustering coefficient $C_i$ of node $i$, \cite{SW} i.e.
\bea
\rho_i=C_i.
\eea

\subsection{Betti numbers}

The neighbourhood topology can be studied by computing the \textit{Betti numbers} of the resulting simplicial complex. These are topological invariants derived from the simplicial complex, and correspond, for each $i \geq 0$, to the number of linearly independent $i$-dimensional holes in the space. Specifically the Betti number $\beta_0$ provides the number of connected components of the neighbourhood, the Betti number $\beta_1$ measures the number of $1$-dimensional holes, i.e. cycles that are not boundaries of $2$-dimensional subsets of the simplicial complex, and so on for spheres, hyperspheres, etc. 

\section{Topology of neighbourhoods in real complex networks}\label{sec:4}

We have considered a number  of large real complex networks, including WWW graphs, social networks and road networks with the total number of nodes $N$, total number of links $L$, average local coefficient $C$, and diameter $D$ indicated in Table \ref{table:tableofvals}. 
All these networks are small-world \cite{SW}, hierarchical \cite{hierarchical} and scale-free \cite{BA}, with the exception of the road networks that are both spatial \cite{Spatial} and planar. For the road networks alone, we have considered neighbourhoods formed by nodes at distance less or equal than $d=5$ from the central nodes. In fact, due to their planarity, the $1$-hop neighbourhoods ($d=1$) are typically formed by isolated nodes. However, for all other small-world networks, considering $5$-hop neighbourhoods would capture non-local properties, since these neighbourhoods would include a large fraction of the nodes of the network. Therefore, for the scale-free, small-world networks (i.e. all the networks considered, with the exception of road networks) we have considered only neighbourhoods formed by nodes at hopping distance $d=1$.
  
For each studied dataset we have performed statistical analysis of the topology of its neighbourhoods by computing the Betti numbers, using the computational homology software CHomP \cite{chomp}. Specifically,  in order to define an efficient computational framework, we have restricted our attention to neighbourhoods formed by clique complexes of dimension equal to $3$, which leads to an accurate information about the Betti numbers $\beta_0$ and $\beta_1$. 
In fact, this procedure guarantees that the value of the measured Betti numbers remain unchanged if clique complexes including simplicies of higher dimensions are also included.

The Betti number $\beta_0$ indicates the number of components of the local neighbourhood. Therefore, a high Betti number $\beta_0$ of a neighbourhood around node $i$ indicates that the node $i$ determining the neighbourhood acts as broker between different otherwise disconnected components of its neighbourhood.
The Betti number $\beta_1$ indicates the number of cycles forming $1$-dimensional holes. Therefore a high ratio $\beta_1/\beta_0$ indicates that the neighbourhood is not maximally dense.

\begin{figure}
     \centering
     \centering
      \includegraphics[width=0.9\columnwidth]{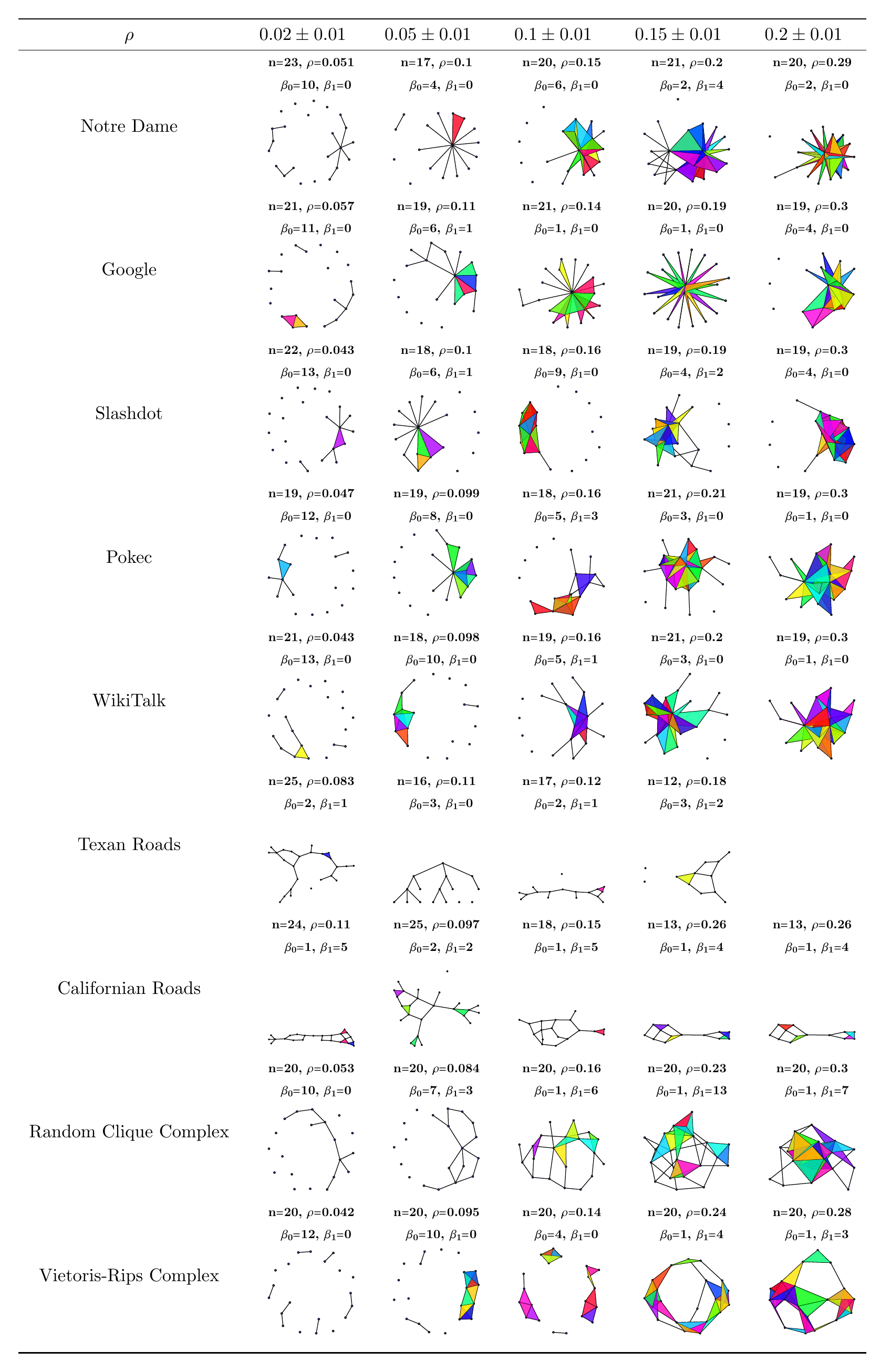}
        \caption{We show several examples of neighbourhoods with number of nodes $n\simeq 20$ and different density of links $\rho$ for different real network datasets and for the random clique complexes.  The topology of network neighbourhoods evaluated by the  Betti numbers $\beta_0$ and $\beta_1$ can have significant fluctuations even for neighbourhoods with comparable values of the local parameters $n$ and $\rho$.}
\label{fig:2}
  \end{figure}

In Figure \ref{fig:2}, we plot several examples of node neighbourhoods found in the analysed datasets.
The first observation that we can draw from the statistical analysis is that if we compare the homology of neighbourhoods with comparable number of nodes $n$ and density of links $\rho$, but coming from different network datasets, there are significant fluctuations. 
 The large variability of the network topology of neighbourhoods with the same density of links $\rho_i$ indicates that the density of links in the neighbourhood (equivalent to the local clustering coefficient $C_i$ for neighbourhood with $d=1$) cannot fully capture the variability observed in the  topology of the node neighbourhoods. In fact for fixed value of the density of links $\rho$ that can have very different Betti numbers $\beta_0$ and $\beta_1$ across different network datasets.
In particular while the small-world and scale-free datasets are characterised often by neighbourhood with high value of the Betti number $\beta_0$, the planar nature of the road network is revealed by the high value of the Betti number $\beta_1$ of its neighbourhoods with respect to the neighbourhoods of the other not planar networks.

\begin{figure}
 \begin{adjustbox}{width=\columnwidth}
     \centering
      \includegraphics[scale=1]{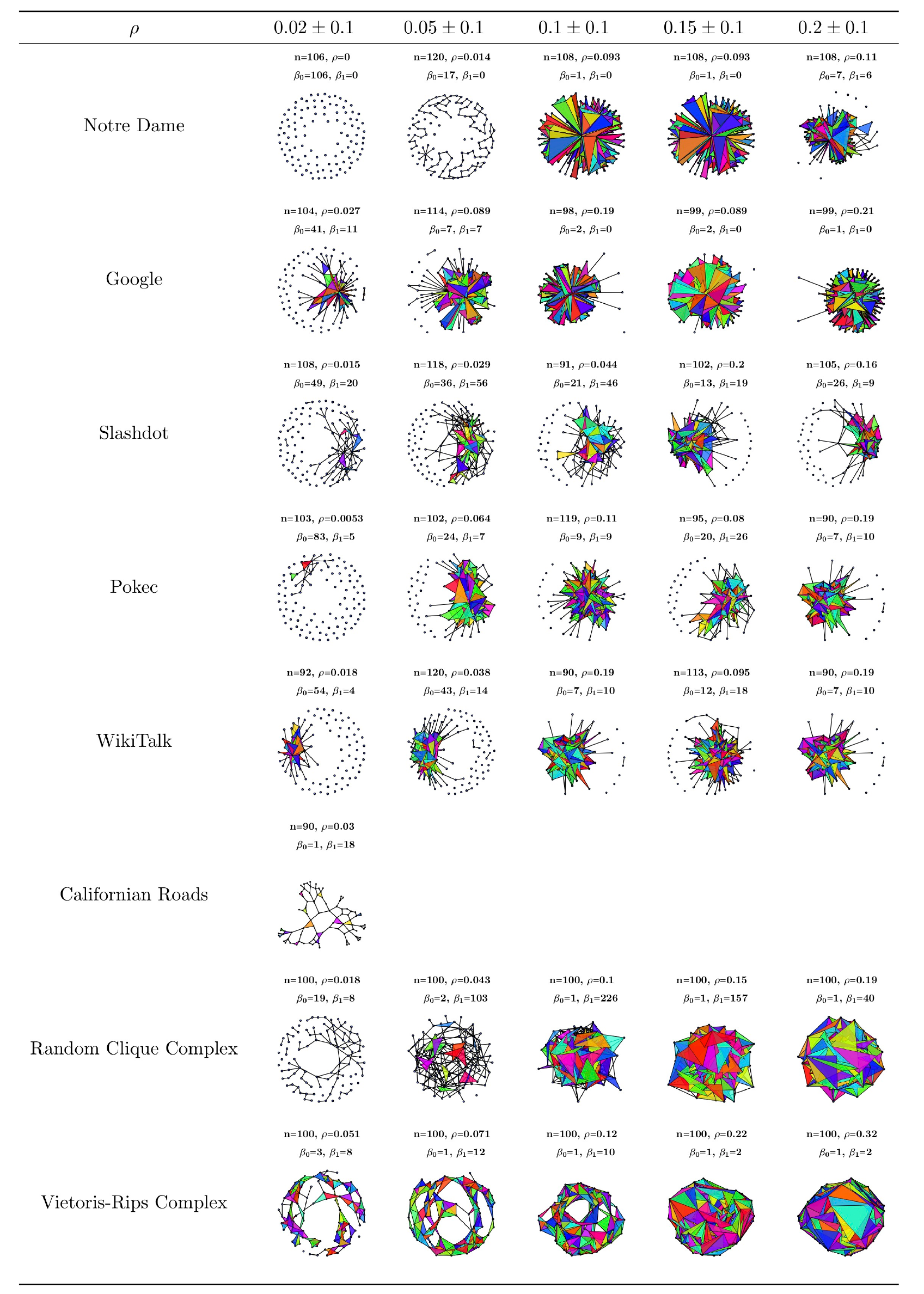}
        \end{adjustbox}
        \caption{We show several examples of neighbourhoods with number of nodes $n\simeq 100$ and different density of links $\rho$ for different real network datasets and for the random clique complexes.  The topology of network neighbourhoods evaluated by the  Betti numbers $\beta_0$ and $\beta_1$ can have significant fluctuations even for neighbourhoods with comparable values of the local parameters $n$ and $\rho$.}
\label{fig:2part2}
  \end{figure}

\begin{table}[tb!]
        \centering
        \begin{tabular}{|c|c|c|c|c|}
        \hline
        \textbf{Network} &  $N$ & $L$ & $C$ &  $D$ \\
        \hline
    Notre Dame Web Graph & 325,729 & 1,497,134 & 0.023  & 46 \\ 
    \hline
    Google Web Graph & 875,713 & 5,105,039 & 0.049  & 21  \\ 
    \hline
     Slashdot Social Network  & 82,168 & 948,464 & 0.060 & 11 \\  
     \hline
    Pokec Social Network  & 1,632,803 & 30,622,564 & 0.010  & 11 \\  
    \hline    
    WikiTalk Social Network  & 2,394,385 & 5,021,410 & 0.010 & 9 \\   
    \hline
    Texas Road Network & 1,379,917 & 1,921,660 & 0.0015  & 1054 \\  
    \hline
    Pensylvania Road Network & 1,088,092 & 1,541,898 & 0.0015 & 786 \\    
    \hline
    California Road Network  & 1,965,206 & 2,766,607 & 0.0015  & 849 \\ 
    \hline
        \end{tabular}
    \caption{Table  of used network datasets including number of nodes $N$, number of links $L$, average clustering coefficient $C$, and Diameter $D$. }
    \label{table:1}
        \label{table:tableofvals}
    \end{table}

\section{Null models of random complexes}\label{sec:5}

\subsection{Why null models?}
In order to characterize real datasets it is always important to refer to suitable null models. A null model allows us to compare the results obtained in real datasets with the results obtained when minimal assumptions are made on the underlying network topology. To this end, here we consider two popular models of simplicial complexes used in the Applied Topology literature, the random clique complex \cite{kahle2009} and the random Vietoris-Rips complex \cite{kahle2}.
It is to be mentioned that this is not the only possible choice of null models. Indeed  several more complex models of stochastic random simplicial complexes \cite{Farber1,Owen1} and evolving simplicial complexes \cite{NGF,Weighted} have been proposed in the literature. However, here the choice of the random clique complex and the random Vietoris-Rips complex is driven by the need to make minimal assumptions on the topology of the node neighbourhoods.

\subsection{Random Clique Complex}

The random clique complex \cite{kahle2009} also known as flag complex is the most fundamental null model for simplicial complexes. The random clique complex is the clique complex of the Erd\"os-R\'enyi random graph $G(n,p)$ of $n$ nodes and density of links $p$.

\subsection{Random Vietoris-Rips  Complex}
The random Vietoris-Rips complex is the clique complex derived from the random geometric graph $\mathcal{G}(n, \pi r_0^2)$ \cite{kahle2}. This ensemble is  formed by $n$ nodes distributed in the unit square $[0,1]^{2}$ with periodic boundary conditions (i.e. $\mathbb{R}^2/\mathbb{Z}^2$)
 according to a Poisson point process and connected when at distance less or equal to $r_0$. Therefore the  expected average degree of the random Vietoris-Rips $1$-skeleton is given by $n \pi r_0^2$. 

\subsection{Betti numbers of random clique complexes and random Vietoris-Rips complexes}

For random clique complexes and random Vietoris-Rips complexes, there are analytical results indicating
the expected value of the Betti numbers and sometimes also their distribution in the limit of large $n$, i.e. for $n\to \infty$ \cite{kahle2009,kahle2}.
In particular, since the Betti number $\beta_0$ of a simplicial complex is  equal to the number of connected components, its value is predicted by the theory of percolation.
For the random clique complex $G(n,p)$ it is well known that in the limit $n\to \infty$ the Betti number $\beta_0$ is a monotone graph property and  is always decreasing with $p$.  In particular for $n\to \infty $ and $np = a\log{n}$ as long as the constant $a$ is  greater than one, the network has almost surely a Betti number $\beta_0=1$, i.e. it is formed by a single connected component. 

The higher Betti numbers  $\beta_j$, with $j>0$ however, display a non-monotone monotonic behaviour with the density of the network $p$. In a random clique complex this implies for instance that the Betti number $\beta_1$ which is zero for $p \ll 1$ initially grows with increasing values of $p$. As the density of links $p$ increases further, $\beta_1$  decreases. This result has an intuitive interpretation. Take for example a 4-cycle, which contributes one to the Betti number $\beta_1$. As the density of links increases, the four nodes on the cycle connect into a four clique, which then contributes zero to the Betti number $\beta_1$.

In Ref. \cite{kahle2009}, Kahle shows how this \textit{uni-modal} transition of the Betti number $\beta_j,j> 0$, which goes from vanishing, to non-vanishing, and back to vanishing again, occurs in the limit $n\to \infty$ in a almost sequential way. For $\alpha,j>0$, if $p = 1/n^{\alpha}$ and $\alpha > 1/j$ or $\alpha < 1/(2j+1)$, then in the limit $n\to \infty$ almost surely we have that the Betti number $\beta_j$ is vanishing, i.e.
\begin{equation}
\beta_j= 0.
\end{equation}
For example, for $j=1$ we have the threshold for percolation, and the theorem states that all components of $G(n,p)$ with $p=1/n^{\alpha}$ and $\alpha>1$ are trees. This is a classic topological property of the component structure in the sub-critical phase of the random graph. The above result implies that for $\alpha < 1/3$, all the cycles belong to the boundary of a higher dimensional clique.

\begin{figure}[!htb]
  \begin{adjustbox}{width=\columnwidth}
    \centering
    \includegraphics[scale=1]{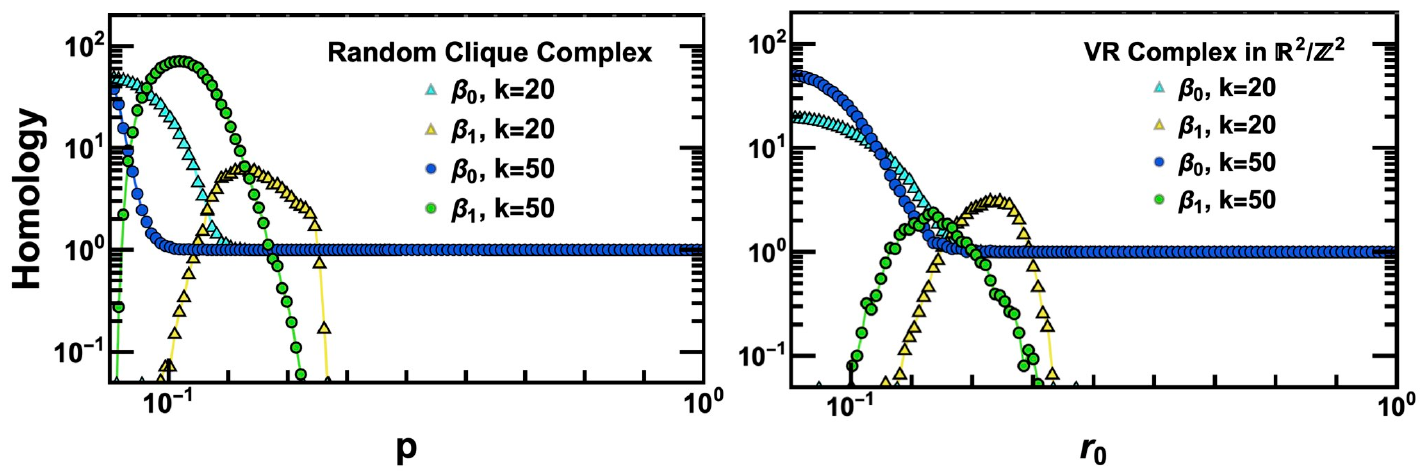}
        \end{adjustbox}
    \caption{ In the left panel  the average Betti numbers $\beta_0$ and $\beta_1$ of the random clique complex of $n=20,50$ nodes are plotted as a function of the density of links $p$. In the right panel  the the average Betti numbers $\beta_0$ and $\beta_1$ of the random Vietoris-Rips complex of $n=20,50$ nodes are plotted as a function of the connection range $r_0$. }
    \label{fig:model}

\end{figure}

Similarly it has been shown \cite{kahle2} that for the random Vietoris-Rips complex, the Betti number $\beta_0$ is monotonically decreasing with the increasing connection range, while the Betti number $\beta_j$ with $j>0$ displays a uni-modal transition with $r_0$.
While these results are clearly fundamental to shed light on the topology of random simplicial complexes, since in this work we consider node neighbourhoods, we need to study how these asymptotic results are reflected in the small or middle sized networks which are available to us.
In  Figure \ref{fig:model}, we show evidence that the discussed asymptotic behaviour of the two chosen null models is also reflected on random clique complexes and random Vietoris-Rips complexes of relatively small size ($n=20$ and $n=50)$.  From this figure, it is apparent that as a consequence of the uni-modality of the Betti number $\beta_1$, typically in a random clique complex with given value of density of links $p$, only one of the two Betti numbers $\beta_0$ and $\beta_1$ are significant. A similar behaviour is observed also for the random Vietoris-Rips complex with given connection range $r_0$.  

\subsection{Comparison between Betti numbers of null models and Betti number of real complex network neighbourhoods}

\begin{figure*}
  \begin{adjustbox}{width=\columnwidth}
     \includegraphics[scale=0.3]{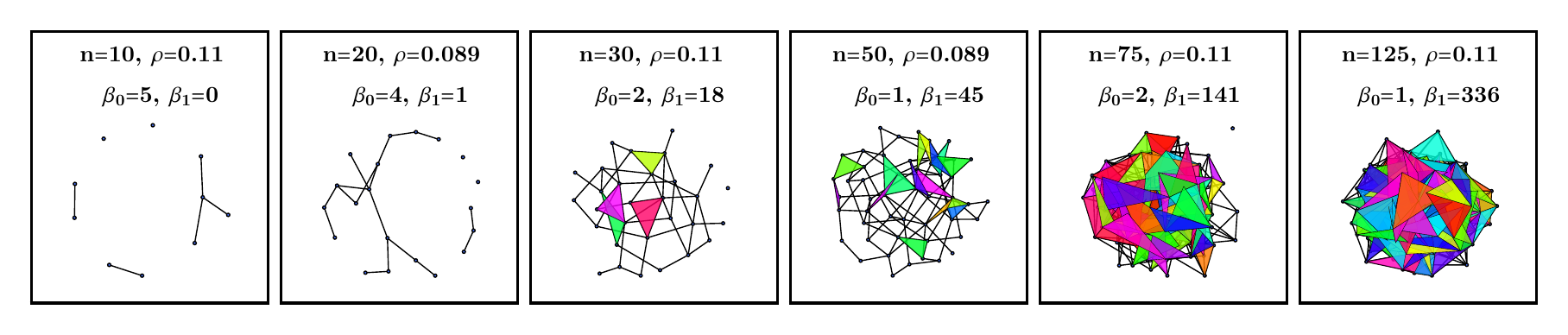}
    \end{adjustbox}
  \caption{The typical neighbouroods of the  random clique complex are shown   for  $\rho\simeq 0.05$, and increasing values of the number of nodes $n=10,20,30,50,75,125$.} \label{fig:a1}
  \vspace{2mm}
  \begin{adjustbox}{width=\columnwidth}
     \includegraphics[scale=0.3]{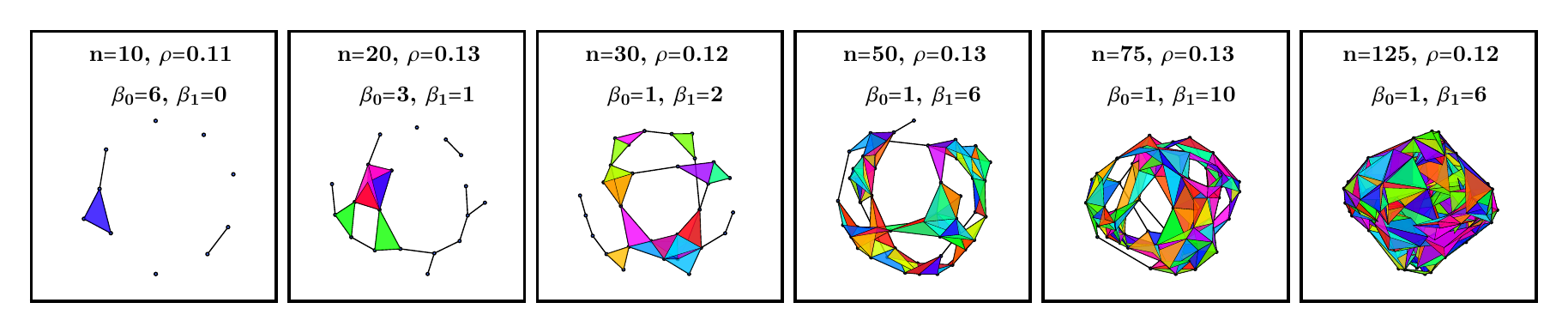}
    \end{adjustbox}
  \caption{The typical neighbourhoods of the  random Vietoris-Rips complex  are shown   for  $\rho\simeq 0.05$ and increasing values of the number of nodes $n=10,20,30,50,75,125$.} \label{fig:a2}
  \end{figure*}
The neighbourhoods of real complex networks datasets can be compared with the two considered null models (random clique complexes and random Vietoris-Rips complexes), whose underlying network skeleton has the same average degree.
When comparing the neighbourhood topology with an ensemble of random complexes (whose members act as an hypothesis of neighbourhood graphs), the neighbourhoods are constructed  according to the null model, rather than obtained by searching large random graphs, as it is done for the real datasets rather random neighbourhoods are directly sampled from the corresponding ensemble of the null models.

\begin{figure*}
  \begin{adjustbox}{width=\columnwidth}
     \includegraphics[scale=0.3]{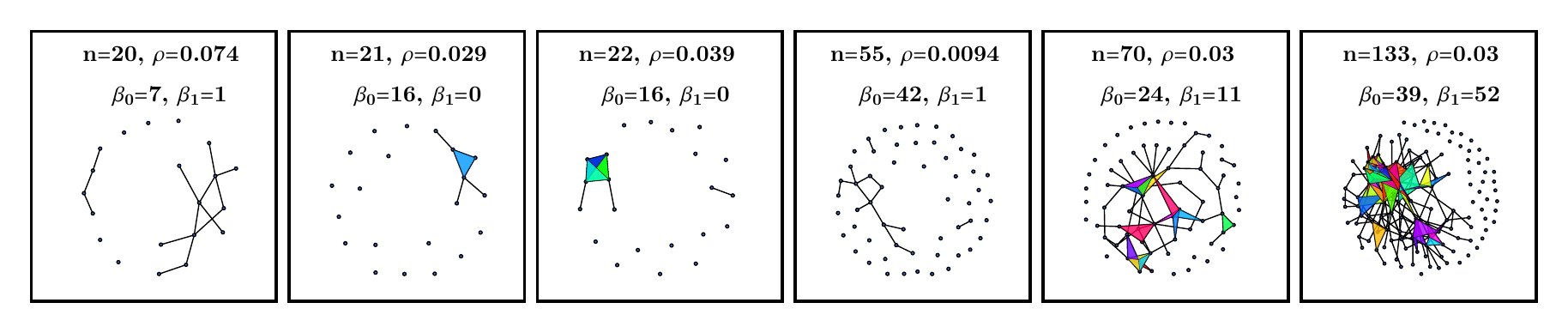}
    \end{adjustbox}
  \caption{A set of typical neighbouroods of the  Slashdot social networks  are shown   for  $\rho\simeq 0.05$ and increasing values of the number of nodes $n=10,20,30,50,75,125$.}\label{fig:a3}
  \vspace{2mm}
  
  \begin{adjustbox}{width=\columnwidth}
     \includegraphics[scale=0.3]{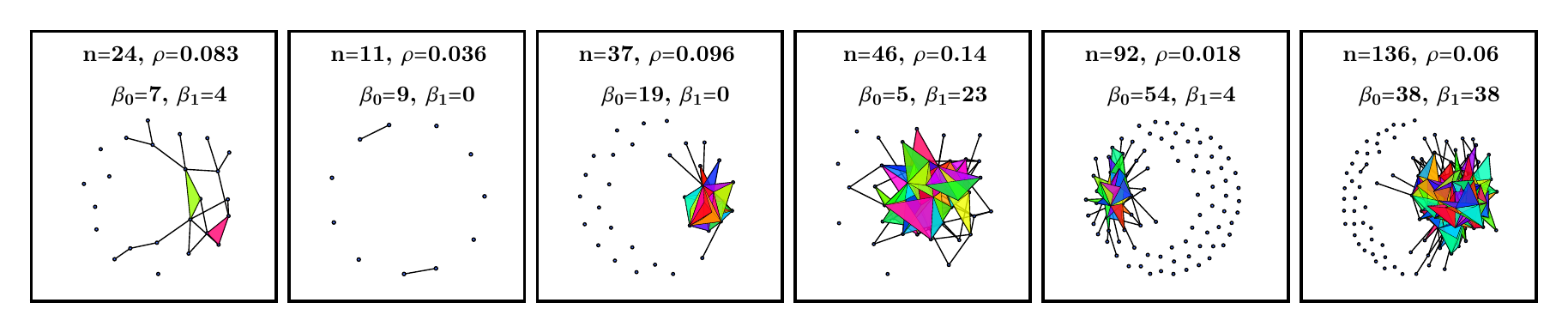}
    \end{adjustbox}
  \caption{A set of typical neighbouroods of the Pokec social network are shown   for  $\rho\simeq 0.05$ and increasing values of the number of nodes $n=10,20,30,50,75,125$.}\label{fig:a4}
    \vspace{2mm}
  
 \begin{adjustbox}{width=\columnwidth}
     \includegraphics[scale=0.3]{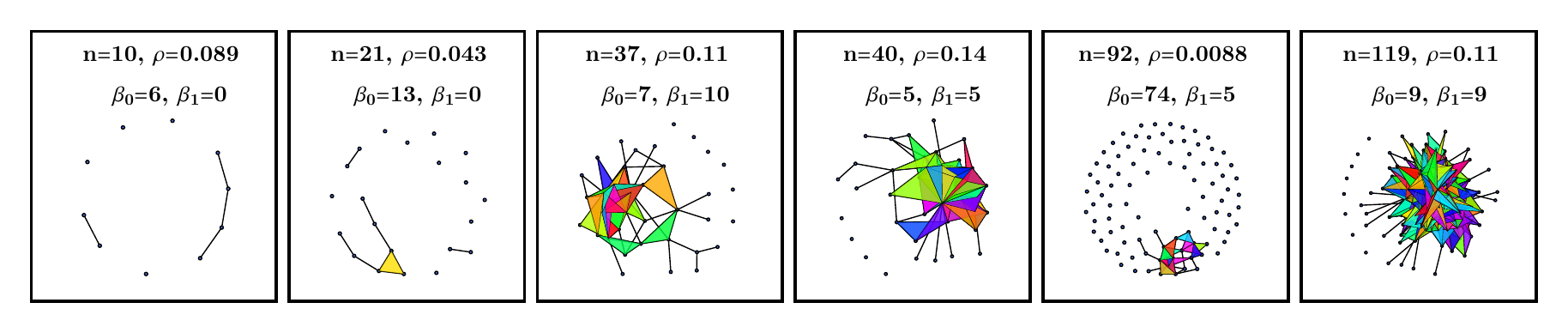}
    \end{adjustbox}
  \caption{A set of typical neighbourhoods of the WikiTalk social network are shown   for  $\rho\simeq 0.05$ and increasing values of the number of nodes $n=10,20,30,50,75,125$.}\label{fig:a5}
  \vspace{2mm}

 \begin{adjustbox}{width=\columnwidth}
     \includegraphics[scale=0.3]{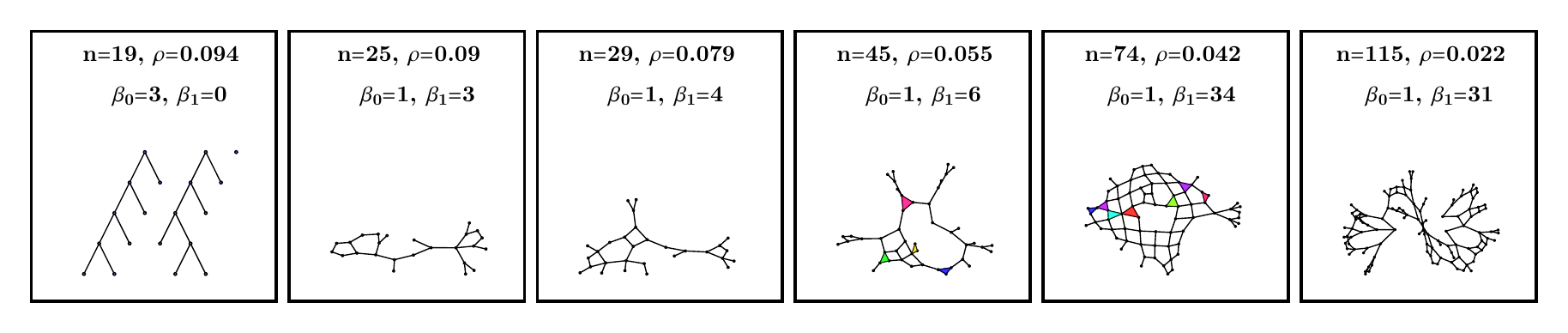}
    \end{adjustbox}
  \caption{A set of typical neighbourhoods of the California road networks are shown   for  $\rho\simeq 0.05$ and increasing values of the number of nodes $n=10,20,30,50,75,125$.}\label{fig:a7}
\end{figure*}

One of the first observation that we draw from this comparison is that the neighbourhoods of the  real network datasets can display at the same time large Betti number $\beta_0$ and large Betti number $\beta_1$ while as we previously discussed, this is in sharp contrast with what it is typically observed in the two considered null models.
In particular in Figures $\ref{fig:a1}$ and $\ref{fig:a2}$ we  show typical realizations of the random clique complexes and the random Vietoris-Rips complexes for given density of links $\rho$ but tuneable size $n$. As the average degree $n\rho$ increases a single connected component quickly forms. We also observe that due to the uni-modal transitions of the Betti number $\beta_1$ the Betti number $\beta_1$ is typically maximal when the Betti number $\beta_0$ is low. 
This behaviour is in sharp contrast to what is observed in real scale-free social network datasets (see Figures $\ref{fig:a3},\ref{fig:a4},\ref{fig:a5}$) tend to have dense neighbourhood formed by several components formed by isolated nodes and a large well clustered component. The observed topology of the node neighbourhoods suggest a local topology of the neighbourhoods similar to the one expected in the  condensed phase of the Strauss model \cite{Strauss,Doro}. This is a model that enforce formation of many triangles leading to a separation between many disconnected isolated nodes and a large, highly clustered component in its condensed phase.
 Additionally in these networks  is not rare to observe neighbourhoods displaying at the same time large Betti number $\beta_0$ and large Betti number $\beta_1$, leading to a topology of network neighbourhoods significantly different from the ones of the considered null models.
In Figure $\ref{fig:a7}$ we show typical instance of neighbourhoods of the  California road network which has instead rather different topology. In fact denser neighbourhoods are typically formed by a single component but display a large Betti number $\beta_1$ reflecting the planar nature of the underlying network.

\section{Statistical topological analysis of complex networks neighbourhoods}\label{sec:6}

\subsection{Homology of hierarchical scale-free networks versus homology of road networks}

The datasets that we have considered include among them several hierarchical scale-free networks \cite{hierarchical} (Notre Dame and  Google Web Graphs,  the Pokec, Slashdot and WikiTalk social networks)
 characterized by an average clustering coefficient of nodes of degree $k$ scaling like
 \bea
 C(k)\simeq k^{-\theta}
 \label{theta}
 \eea
 and several road networks (Texas, Pensylvania, California road networks).

   \begin{figure*}
     \noindent \begin{centering}
       \begin{adjustbox}{width=\columnwidth}
         \includegraphics[scale=0.8]{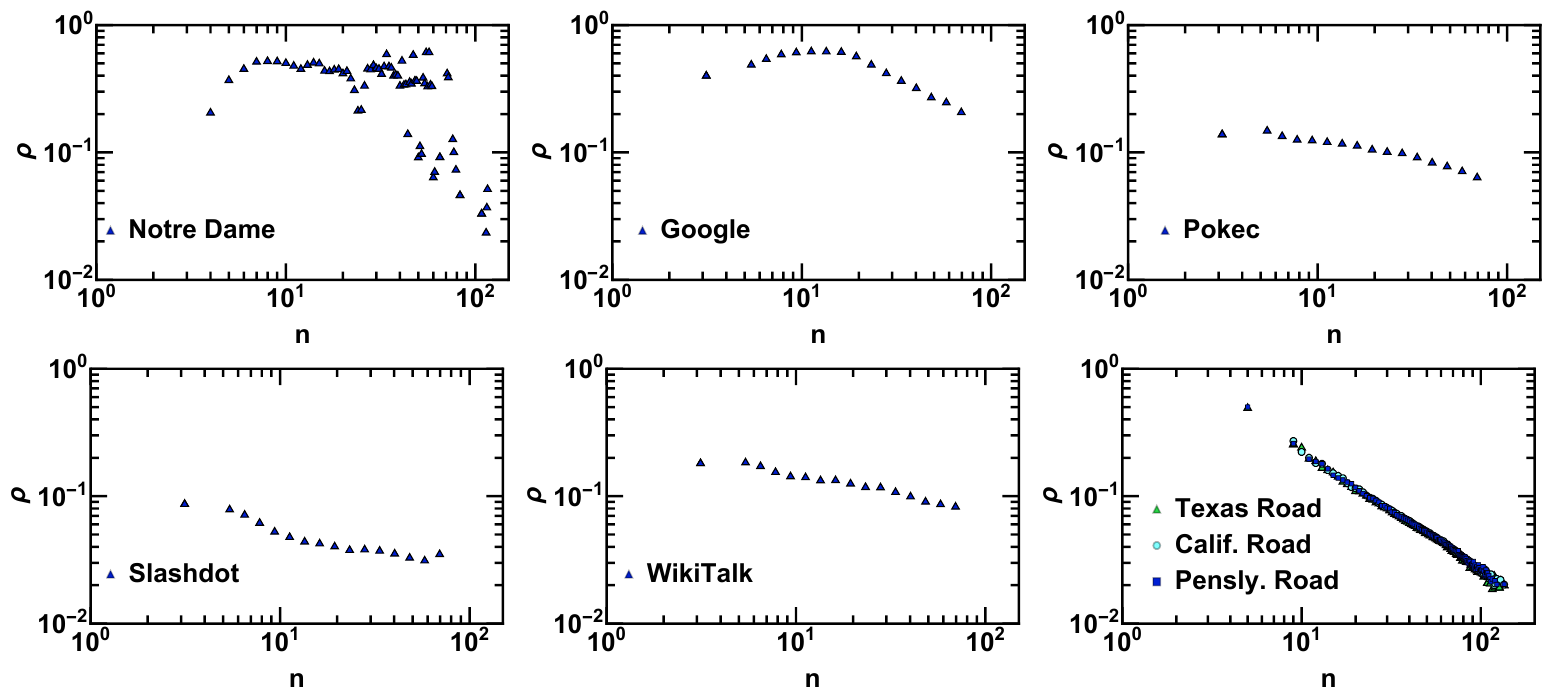}
         \end{adjustbox}
    \caption{The average density of links $\rho$ of  node neighbourhoods of size $n$ is plotted as a function of $n$  different network datasets.}
    \label{fig:rcc0}
 \par\end{centering}
 \noindent \begin{centering}
   
   \begin{adjustbox}{width=\columnwidth}
     \includegraphics[scale=0.8]{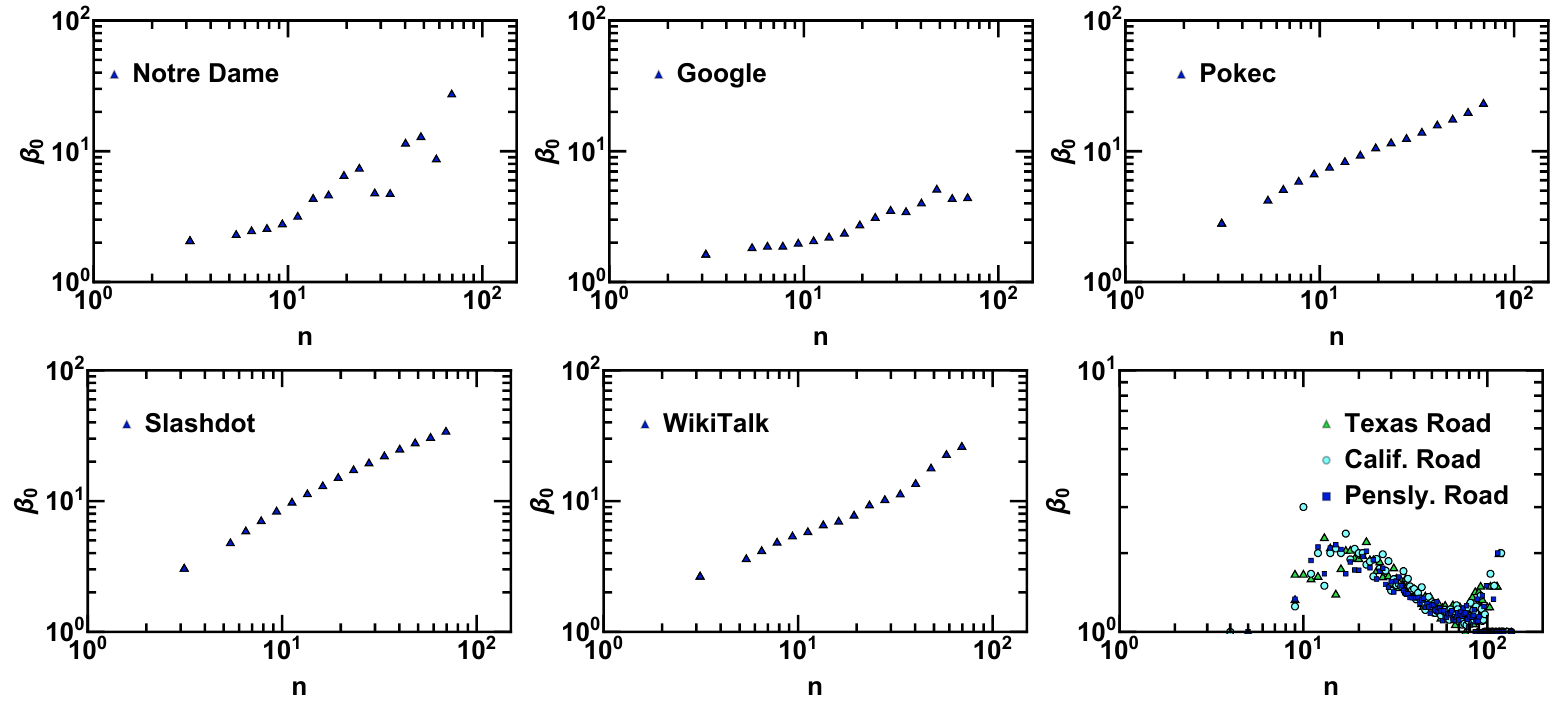}
     \end{adjustbox}
    \caption{The average Betti number $\beta_0$ of node neighbourhodds of size  $n$ is plotted as a function of $n$ for different network datasets.}
    \label{fig:rcc}
 \par\end{centering}
  \end{figure*}

  \begin{figure*}
   \noindent \begin{centering}
     \begin{adjustbox}{width=\columnwidth}
       \includegraphics[scale=0.8]{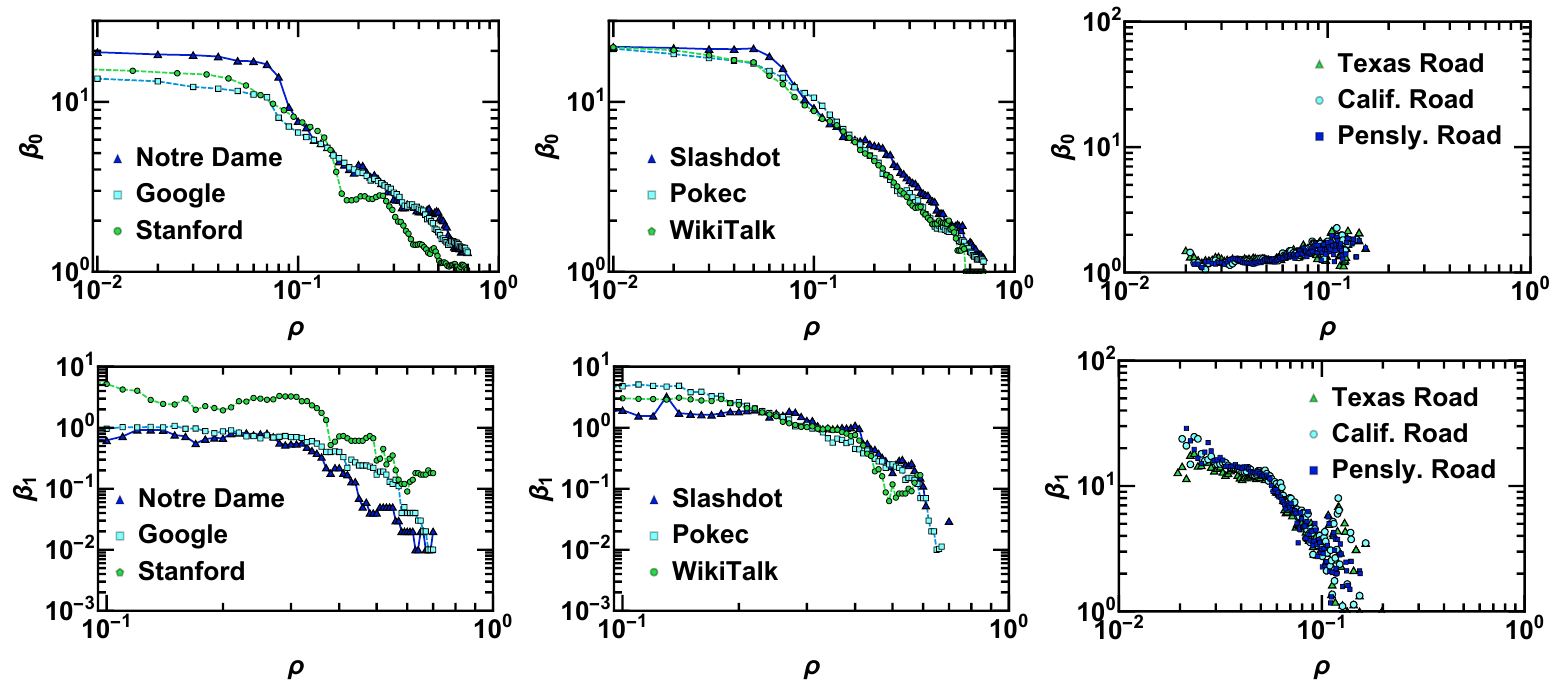}
     \end{adjustbox}
     \caption{The average Betti numbers $\beta_0$ and $\beta_1$  of node neighbourhoods with given  density of links $\rho$ are plotted as a function of $\rho$ for different network datasets.}\label{fig:b0C1}
 \par\end{centering}
  \end{figure*}

 Clearly we expect to see significant differences in the topology of the neighbourhoods of hierarchical networks and road networks. The planarity of road networks clearly constraint all the Betti numbers $\beta_j$ with $j>1$ to be zero. However as we have seen in the typical neighbourhood of road networks shown in Figure $\ref{fig:2}$ and in Figure $\ref{fig:a7}$, the road networks neighbourhood tends also to have a lower $\beta_0$ and a higher $\beta_1$ with respect to neighbourhoods in the other datasets having the same number of nodes $n$ and density of links $\rho$.
 To investigate further the statistical differences between road network neighbourhoods and scale-free hierarchical networks neighbourhoods we have evaluated the average value of the Betti numbers $\beta_0$ and $\beta_1$ over neighbourhoods with fixed number of nodes $n$ or density of links $\rho$. 
In the hierarchical scale-free   networks a node with  degree $k$ will give rise to a node neighbourhood of $n=k$ nodes as long as all the nodes of the neighbourhood are at distance $\delta=1$ from the original node. Correspondingly the density of links $\rho$ of the node neighbourhood can be approximated with the average clustering coefficient $C(k)$ of nodes of degree $k=n$, i.e. $\rho \simeq C(k)$ that obeys Eq. ($\ref{theta}$) providing the following power-law scaling  of $\rho$ as a function of $n$ (see Figure $\ref{fig:rcc0}$)
\bea
\rho \propto n^{-\theta}.
\label{rhon}
\eea
From the statistical topological analysis of the  average Betti number $\beta_0$ of the network neighbourhood we find that the average Betti number $\beta_0$ increases as a power law with $n$, i.e.
\begin{equation}
  \beta_0 \propto n^{\nu}, 
\end{equation}
with $\nu>0$ (see  Figure $\ref{fig:rcc}$).
This scaling indicates a monotonic power-law increases of the number of components of the local neighbourhoods as a function of the number of nodes $n$ of the neighbourhoods. Therefore hubs tend to be broker between different otherwise unconnected communities.
In  Figure  $\ref{fig:b0C1}$ we report the average Betti numbers $\beta_0$ and $\beta_1$ for node neighbourhoods with given density of links $\rho$.
For hierarchical scale-free networks, using the  scaling indicated in Eq.$(\ref{rhon})$, it is easy to predict that the average Betti number $\beta_0$ should decay as an inverse power of the density of links $\rho$, i.e.
\begin{equation}
  \beta_0\propto \rho^{-\alpha}
  \label{alpha}
\end{equation}
with $\alpha=\nu/\theta$. 
These scaling relations imply that more densely connected neighbourhoods are typically smaller, and characterised by a smaller Betti number $\beta_0$. 
The road networks, that are not hierarchical are characterised by a significant different trend of the average Betti number $\beta_0$ as a function of $\rho$. In fact  for the road networks the  Betti number $\beta_0$  is  a non-decreasing function of $\rho$. 
This result reveal that the number of connected components of the road neighbourhoods decreases for the neighbourhood of  larger  road junctions.

Therefore the Betti number $\beta_0$ of road networks neighbourhoods display very relevant statistical differences with respect to the Betti number $\beta_0$ of scale-free hierarchical networks.
\begin{figure*}
  \noindent \begin{centering}
    \begin{adjustbox}{width=\columnwidth}
      \includegraphics[scale=0.8]{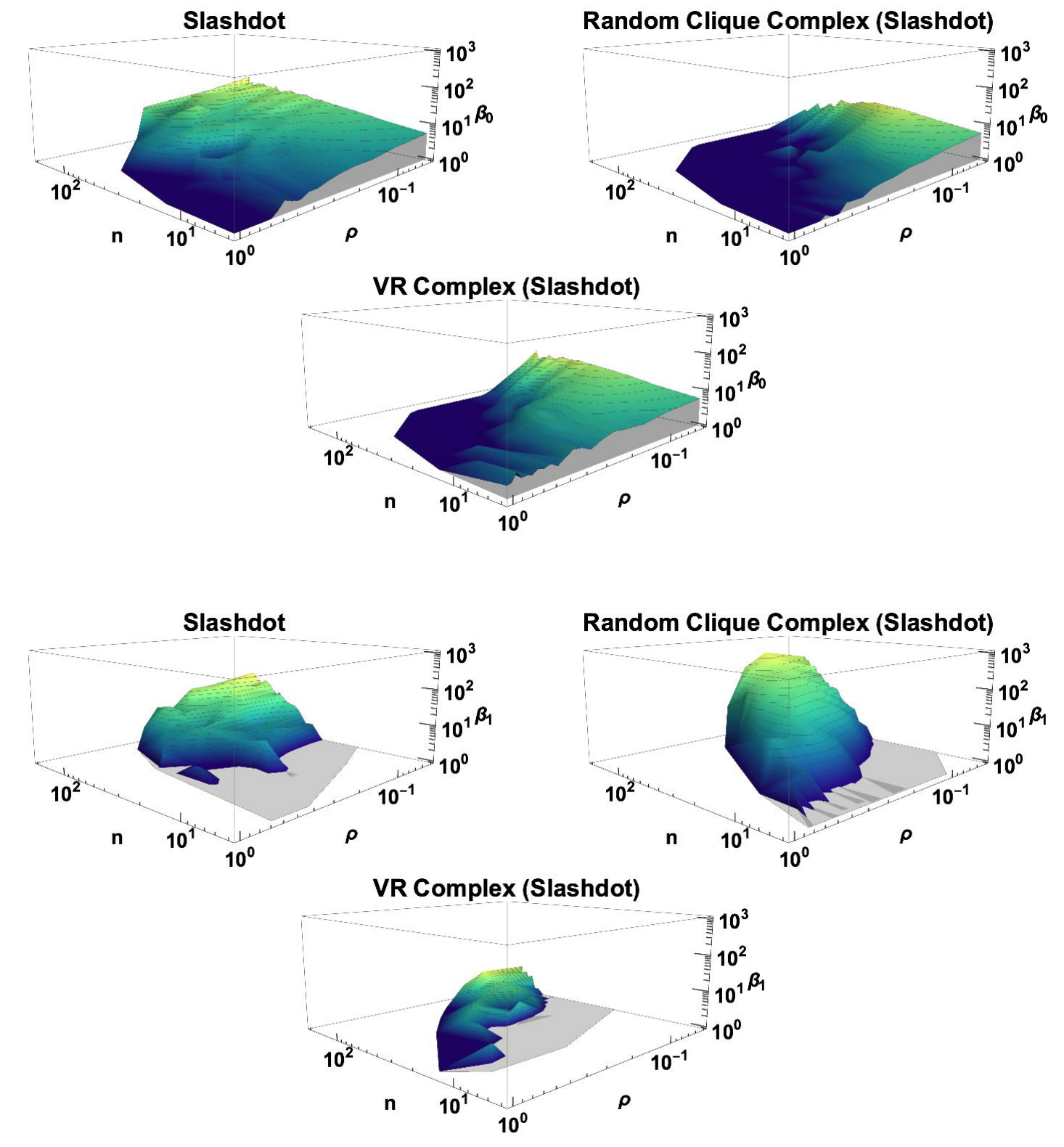}
      \end{adjustbox}
    \caption{The average Betti numbers $\beta_0$ and $\beta_1$ of the neighbourhoods to the   Slashdot social network as plotted as a function of  the number of nodes $n$ and the density of links $\rho$ of the neighbouroods.  The real data are then compared with the results obtained on the random clique complex and the random Vietoris-Rips complex observing significant difference highlighting the non-random character of the real dataset.}\label{fig:3dslash}
    \par\end{centering}
\end{figure*}

\begin{figure*}
  \noindent \begin{centering}
    \begin{adjustbox}{width=\columnwidth}
      \includegraphics[scale=0.8]{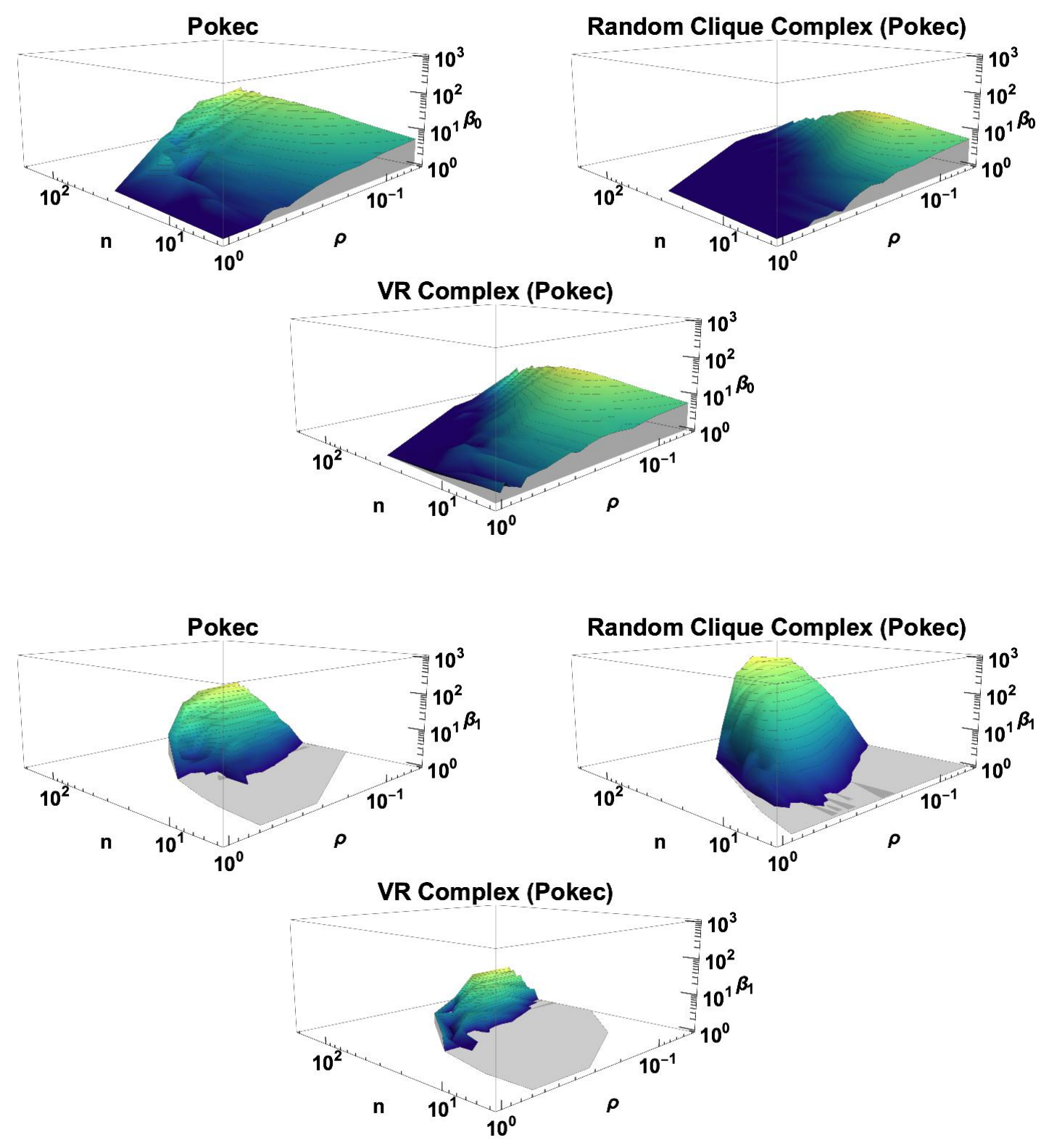}
      \end{adjustbox}
    \caption{The average Betti numbers $\beta_0$ and $\beta_1$ of the neighbourhoods to the  Pokec social network as plotted as a function of  the number of nodes $n$ and the density of links $\rho$ of the neighbourhoods.  The real data are then compared with the results obtained on the random clique complex and the random Vietoris-Rips complex observing significant difference highlighting the non-random character of the real dataset.}\label{fig:3dpokec}
    \par\end{centering}
\end{figure*}

\subsection{Topological data analysis as a function of the number of nodes $n$ and the density of links $\rho$}

The non-random behaviour in the datasets becomes apparent when we characterise the average topology of the neighbourhoods of real datasets with given number of nodes $n$ and density of links $\rho$. To this end we plot the average Betti numbers $\beta_0$ and $\beta_1$ as a function of $n$ and $\rho$ and we compare the results with the average Betti numbers of the random clique complexes and the random Vietoris-Rips complexes with the same number of nodes and density of links. 
The Figures \ref{fig:3dslash} and \ref{fig:3dpokec} display the numerical results for the neighbourhoods of the Slashdot social network and of the Pokec social network, respectively.

\begin{figure*}
  \noindent \begin{centering}
    \begin{adjustbox}{width=\columnwidth}
      \includegraphics[width=0.9\columnwidth]{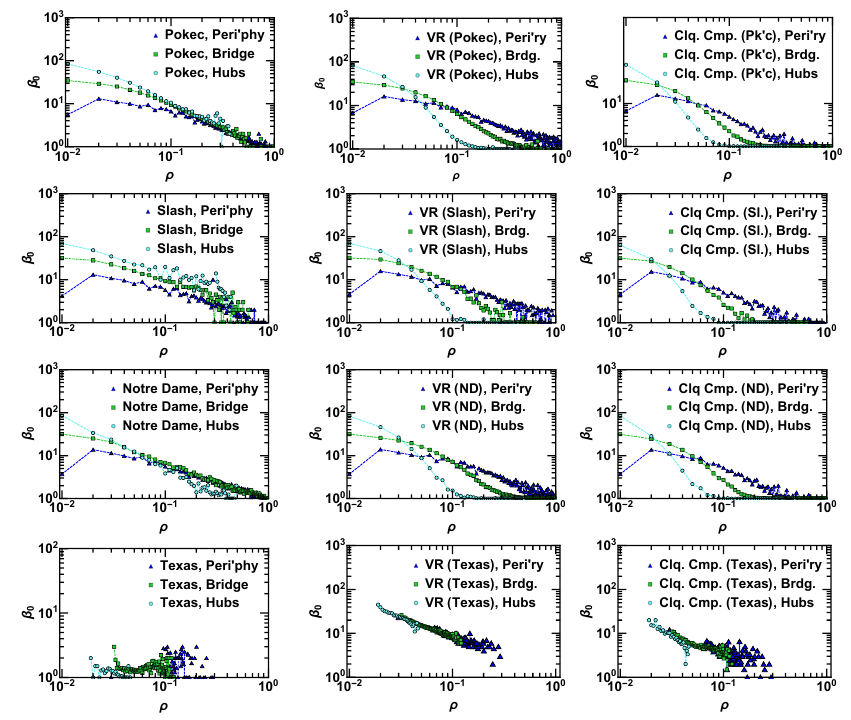}
      \end{adjustbox}
    \caption{The average Betti number $\beta_0$ of neighbourhoods of the Slashdot and  Pokec social networks, of the Notre Dame Web graph  and of the Texas Roads Networks, are plotted as a function of the  density of links $\rho$ by distinguishing between neighbouroods of different size $n$ ($n<20$, $20\leq n\leq 70$ and $n>70$). The results obtained with the real datasets are compared with the results obtained for the random clique complex and the random Vietoris-Rips complex.}
    \label{fig:ndep}
    \par\end{centering}
\end{figure*}

In order to emphasise the quantitative differences that we observe in the real datasets with respect to the null models in Figure \ref{fig:ndep} we plot the average Betti number $\beta_0$ of neighbourhoods of hubs nodes ($n>70$), peripheral nodes $(n<20)$ and of bridge nodes $n\in [20,70]$ as a function of $\rho$.
All the considered real scale-free networks display a power law dependence of the Betti number $\beta_0$ with respect to the density of links $\rho$ that persists throughout all instances, independently of $n$. However, this behaviour is not captured by the null models. This is the clearest distinction between the stochastic topology of the null models, and the neighbourhoods in the real datasets.

This apparently follows from the scale-free, hierarchical structure of the datasets, which acts to ensure that large neighbourhoods display multiple disconnected components, unlike  random cliques or random Vietoris Rips complexes.
The Texas road network datasets, also, display a topology that is significantly different from the null models, where the Betti number $\beta_0$ is not decreasing with the density of links $\rho$ while it is decreasing for the random clique complex and for the random Vietoris-Rips complex.

\section{Conclusions}\label{sec:7}
In this paper, we have analysed the topology of node neighbourhoods in large network datasets. The node neighbourhood of a generic node $i$ is the clique complex of the network induced by the nodes up to distance $d$ from $i$. The topology of node neighbourhoods is then investigated by calculating their Betti numbers $\beta_0$ and $\beta_1$. A node neighbourhood with a high $\beta_0$ indicates that the central node is connected to many nodes that are not directly connected to any other node in the neighbourhood. A high $\beta_1$ indicates instead that among connected nodes in the neighbourhood, there are several open cycles, implying that the corresponding cluster is not densely connected.

Our large scale statistical analysis reveals that the topology of these neighbourhoods is not only determined by their size and their link density. In fact, for a given size and link density of a neighbourhood, different topologies can be observed. Specifically we show that the topological study of node neighbourhoods is able to distinguish between the neighbourhoods of scale-free hierarchical networks and the neighbourhoods of spatial road networks. Moreover both types of real datasets obey significant organisation principles that impose a local topology of the node neighbourhoods that is significantly different from the random clique complex and the random Vietoris-Rips complex.
In the future an interesting question that we would like to investigate is to what extent the recently proposed local curvatures of discrete networks and simplicial complexes \cite{Jost1,Jost2,Loll} capture the local topology of node neighbourhoods.

\end{document}